\documentclass[prb,twocolumn,floats,floatfix,superscriptaddress,
               showpacs,amssymb,amsmath,amsfonts]{revtex4}%
\usepackage{graphicx}
\usepackage{dcolumn}
\usepackage{bm}

\begin{document}
\title{Model Hamiltonian parameters for half-metallic ferromagnets 
NiMnSb and CrO$_2$} 

\author{A. Yamasaki}
\email{A.Yamasaki@fkf.mpg.de}
\affiliation{Max-Planck-Institut f\"ur Festk\"orperforschung
Heisenbergstrasse 1, D-70569 Stuttgart, Germany} 

\author{L. Chioncel}
\affiliation{Department of Physics, University of Oradea, 410087 Oradea, 
             Romania}
\affiliation{Institute for Theoretical Physics and Computational
Physics, Graz University of Technology, A-8010 Graz, Austria}

\author{A. I. Lichtenstein}
\affiliation{Institute of Theoretical Physics, University of Hamburg, Germany}

\author{O. K. Andersen}
\affiliation{Max-Planck-Institut f\"ur Festk\"orperforschung
Heisenbergstrasse 1, D-70569 Stuttgart, Germany} 

\date{Received 10 March 2006; revised manuscript received 25 May 2006}

\begin{abstract}
Using the recently developed $N$th-order muffin-tin-orbital (NMTO) based 
downfolding technique we revisit the electronic properties of half-metallic 
ferromagnets, the semi-Heusler NiMnSb and rutile CrO$_2$. The 
NMTO Wannier orbitals for the Mn-{\it d} and Cr-{\it $t_{2g}$} manifolds 
are constructed and the mechanism of chemical bonding is discussed. The 
effective hopping Hamiltonian parameters are calculated using a NMTO 
downfolded basis set. We propose model Hamiltonian parameters with 
possibly minimal basis sets for both half-metallic ferromagnetic alloys.
\end{abstract}


\maketitle

\section{Introduction}

Half-metallic ferromagnets (HMF) such as the semi-Heusler NiMnSb or 
the rutile CrO$_2$ are a subject of strongly growing interest mainly 
due to their potential applications in spin-dependent 
electronics, so-called ``spintoronics.''~\cite{ufn,Wolf01,sarma1,deGroot83}
Concerning their electronic structure, half-metallic materials exhibit 
a gap in one spin channel and a normal metallic behavior for the opposite
spin bands. This means that electrons at the Fermi level are 100\% 
spin polarized. Much experimental effort is devoted to the semi-Heusler 
NiMnSb compound, although highly spin-polarized carrier injection 
has not yet been achieved. 
At low temperatures superconducting point contact measurements revealed
a less then $50\%$ polarization of valence electrons.~\cite{Soulen98,Clowes04}
Similar values were obtained by spin resolved 
photoemission,~\cite{Zhu01} and more recently experiments using the 
synchrotron radiation showed at room temperature a 40\% polarization at the 
Fermi level.~\cite{Sicot06} This discrepancy in comparison with the 
predicted half-metallic behavior is attributed to interface and surface 
effects. 
However, by 
a proper engineering, half-metalicity can be restored at 
surface/interfaces.~\cite{Wijs01} 
The experimental evidence for half-metallicity is stronger in 
CrO$_2$: Andreev reflection,~\cite{Ji01} superconducting tunneling,~\cite{Parker01}
photoemission,~\cite{Dedkov02} and point-contact magnetoresistance,~\cite{Coey02} 
all give values of polarization in the range 85--100 \%.  
Keizer {\it et al.} have injected new excitement into
the field of half metals, by reporting the existence of a spin triplet 
supercurrent through the strong ferromagnet CrO$_2$.~\cite{Keizer06}

Future spin electronic devices based on HMF will probably be expected 
to work around and above room temperature, so one of the essential
requirements is that these ferromagnets should have quite high Curie 
temperatures. Both NiMnSb ($T_c$=730~K) and CrO$_2$ ($T_c$=400~K) fulfill 
this requirement. A second essential requirement for using half-metallic 
materials in practical devices could be the understanding of finite-temperature 
behavior of spin polarization both from the experimental and the theoretical point 
of view. 

On the theoretical side, HMF have been strongly supported by first-principles 
calculations, based on density-functional theory.~\cite{deGroot83} These calculations
offer a proper description of the ground state properties, and are usually performed
for zero temperatures. One way to approach the finite temperature behavior is 
based on modeling the many-body interactions in the real material, therefore the 
evaluation of the model Hamiltonian parameters is required. These parameters 
constitute the starting point for a finite-temperature 
full many-body microscopic description. The local density approximation (LDA) 
electronic structure allows us to evaluate the effective hopping parameters, to which 
a Hubbard type interaction is added to construct the starting Hamiltonian. In the 
multiorbital case the Hubbard Hamiltonian is described by 
\begin{align}
\hat{H}^{\rm int}=&\frac{1}{2}\sum_{imm',\sigma}U_{mm'}
\hat{n}_{im\sigma}\hat{n}_{im'-\sigma} \notag \\
&+\frac{1}{2}\sum_{imm'(\neq m),\sigma}(U_{mm'}-J_{mm'})
\hat{n}_{im\sigma}\hat{n}_{im'\sigma},
\label{hub_ham}
\end{align}
where $\sigma$ is the spin index, $m,m'$ are local orbitals at site $i$. The on-site 
Coulomb interactions are expressed in terms of two parameters: 
$U_{mm}=U$, $U_{mm'(\neq m)}=U-2J$, and $J_{mm'}=J$.~\cite{ourDMFT} 
This model Hamiltonian is defined on a basis set of local 
orbitals and thus the microscopic interactions are local, involving a small number of 
electrons and a small number of orbitals. A numerically exact solution of this model 
is achieved by the quantum Monte Carlo solver in the framework of the dynamical mean 
field theory (DMFT).~\cite{DMFT1,DMFT2} 
We used the recently developed LDA+DMFT 
methods~\cite{anisDMFT,ourDMFT,Katsnelson99,HeldLDADMFT,EMTODMFT,today} 
in order to investigate finite-temperature 
many-body effects for the practically important spintronic materials 
such as the semi-Heuslers 
NiMnSb,~\cite{Chioncel03} FeMnSb,~\cite{femnsb} 
and the zinc-blende CrAs (Ref.~\onlinecite{Chioncel05}) and VAs.~\cite{vas}
In these materials, due to their
half-metallic ferromagnetic band structure, the incoherent 
[nonquasiaprticle (NQP)] states play an important role. 
The NQP states were considered theoretically for the first time 
by Edwards and Hertz~\cite{edwards} in the framework of broad-band Hubbard model 
for itinerant electron ferromagnets.

For the realistic electronic structure 
of NiMnSb NQP states are situated just above the Fermi level for the minority 
spin channel, having a considerable spectral weight.~\cite{Chioncel03} In FeMnSb
the spectral weight of NQP states is enhanced in comparison with NiMnSb, producing 
a drastic depolarization at the Fermi level.~\cite{femnsb} For CrAs in the zinc-blende 
structure,~\cite{Chioncel05} the spectral weight of NQP states was studied in connection 
with the substrate lattice parameter. For large substrate lattice parameters the Fermi level
is situated close to the middle of the minority spin gap and the NQP states are clearly 
visible. However, for smaller substrate lattice parameters the NQP contribution is  
negligible.  VAs in a similar zinc-blende structure is predicted by LDA/GGA to be 
a narrow gap semiconductor.~\cite{Sanyal} In addition to the presence of NQP states, 
many-body interactions determine the closure of the semiconducting majority spin gap, 
leading to a half-metallic ferromagnetic ground state.~\cite{vas} Therefore, a correct 
prediction of new spintronic materials should take into account finite-temperature 
many-body correlation effects, which might play an essential role in depolarization. 

In order to further investigate the nature of the NQP state, 
a realistic model Hamiltonian 
is required. In this paper we will use the recently developed massive downfolding 
scheme~\cite{andersen00,zurek05} 
in order to produce real space Hamiltonian parameters for both 
NiMnSb and CrO$_2$ half-metallic compounds. 

The paper is organized as follows. Section \ref{eff_par} describes briefly the 
computational details of the $N$th-order muffin-tin orbital (NMTO) method. 
The corresponding subsections \ref{downnimnsb}
and \ref{downcro2} present the results of downfolding onto the Mn-{\it d},  
Cr-$t_{2g}$ manifolds, respectively, and gives the values of the effective hopping parameters. 
In the case of CrO$_2$ we compare the results of the matrix elements of the 
effective hopping Hamiltonian in two distinct Wannier orbitals basis sets: the one 
which describes the full Cr-$t_{2g}$ manifold and a second in which $d_{xy}$ and 
$d_{yz\pm zx}$ orbitals could be described individually. 
The values of the effective Coulomb interaction parameters are given in Sec. \ref{effU}.
In the summary we discuss the construction of possible model Hamiltonians with a minimal basis set
for both half-metallic ferromagnets.

\begin{table}[t]
\caption{\label{tab1}
The muffin-tin radii for NiMnSb and CrO$_2$. 
The radii of the empty spheres $E$ are also given. 
The second raw 
coresponds to the LMTO basis sets used in the self-consistent calculation
of LDA potential. 
$(l)$ means that the $l$-partial waves were downfolded within the LMTO-ASA
+ cc. 
}
\begin{ruledtabular}
\begin{tabular}{ccccccc}
&               & \multicolumn{4}{c}{NiMnSb}      \\
&               & Ni    & Mn     & Sb     & $E$   \\ \cline{3-6}
&$R_{\rm MT}$ (a.u.) & 2.584 & 2.840  & 2.981  & 2.583 \\ 
&basis set & $spd$ & $spd$ & $sp$($df$) & $sp$($d$) \\ \\
&               & \multicolumn{4}{c}{CrO$_2$}     \\
&               & Cr    & O      & $E$    & $E1$  \\ \cline{3-6}
&$R_{\rm MT}$ (a.u.) & 2.213 & 2.094  & 1.653  & 1.566 \\
&basis set & $spd$ & ($s$)$p$($d$) & $s$($p$)  & $s$($p$)      
\end{tabular}
\end{ruledtabular}
\end{table}

\begin{figure}[b]
\includegraphics[width=\columnwidth]{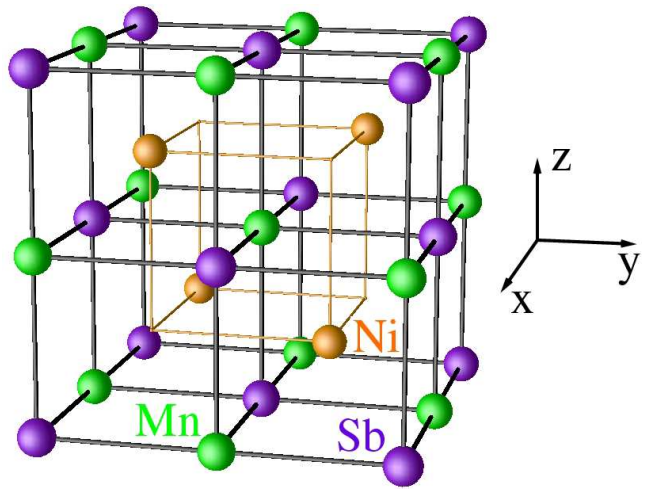} 
\caption{(Color online) $C$1$_b$ structure with the fcc Bravais lattice 
(space group $F\overline{4}3m$).
Mn (green) and Sb (purple) atoms are located at (0, 0, 0) and 
($\frac{1}{2}$, $\frac{1}{2}$, $\frac{1}{2}$) forming the rocksalt structure
arrangement.
Ni (orange) atom is located in the octahedrally coordinated pocket, 
at one of the cube
center positions ($\frac{3}{4}$, $\frac{3}{4}$, $\frac{3}{4}$) leaving
the other ($\frac{1}{4}$, $\frac{1}{4}$, $\frac{1}{4}$) empty.
This creates voids in the structure.
} 
\label{NiMnSbstruc}
\end{figure}

\section{Effective hopping parameters\label{eff_par}}
In the present paper we use the NMTO
method~\cite{andersen00,zurek05} for generation of localized Wannier functions.
The NMTO method can be used to generate truly minimal basis sets with massive 
downfolding technique. Downfolding produces minimal bands which follows 
exactly the bands obtained with the large basis set. In the case of NiMnSb 
and CrO$_2$, Mn-$d$ and Cr-$t_{2g}$ form the minimal basis set. 
The truly minimal set of symmetrically orthonormalized NMTOs is a set of Wannier 
functions. In the construction of the NMTO basis set the active channels are forced 
to be localized onto the eigenchannel ${\bf R}lm$, therefore the NMTO basis set 
is strongly localized.

Fourier transformation of the orthonormalized NMTO Hamiltonian, 
$H^{\rm LDA}({\bf k})$, yields on-site energies and hopping integrals,
\begin{equation}
H^{\rm LDA}_{{\bf 0}m',{\bf R}m} \equiv \left\langle \chi _{%
{\bf 0}m^{\prime }}^{\perp }\left\vert \mathcal{H}%
^{\rm LDA}-\varepsilon _{F}\right\vert \chi _{{\bf R}m}^{\perp}\right\rangle 
\equiv t_{m^{\prime },m}^{xyz},
\end{equation}
in a Wannier representation, where the NMTO Wannier function
$\left\vert \chi _{\mathbf{R}m}^{\perp}\right\rangle$ is orthonormal.

 The matrix element between orbitals $m^{\prime }$ and 
$m$, both on site ${\bf R}'$=${\bf R}$=${\bf 0}$, is $t_{m^{\prime},m}^{\bf 0}$, 
and the hopping integral from orbital $m^{\prime }$ on site 
${\bf R}'$=${\bf 0}$ to orbital $m$ on site ${\bf R}$=($x,y,z$) is 
$t_{m^{\prime },m}^{xyz}$.

The LDA potential is generated with the Stuttgart TB-LMTO-ASA code (the LMTO-ASA  
including the combined correction).~\cite{lmto,tblmto,tb47} NMTO calculations 
are performed using the generated LDA potentials. The radii of MT potential spheres 
and the LMTO bases used in the calculation for NiMnSb and CrO$_2$ are 
listed in Table~\ref{tab1}. For the detail of the calculation, see the appendixes of 
Ref.~[\onlinecite{PavariniNJP}].

\subsection{Downfolding onto the Mn-{\it d} manifolds in NiMnSb\label{downnimnsb}}

The intermetallic compound NiMnSb crystallizes in the cubic structure of MgAgAs 
type ($C1_b$) with the fcc Bravais lattice (space group $F\overline{4}3m=T_d^2$).
The crystal structure is shown in Fig.~\ref{NiMnSbstruc}. 
This structure can be described as three interpenetrating fcc lattices of Ni, Mn, 
and Sb. The Ni and Sb sublattices are shifted relative to the Mn 
sublattice by a quarter of the $[111]$ diagonal in opposite directions. In the present 
calculation the experimental lattice constant of NiMnSb ($a$=5.927 \AA) is used.

A detailed description of the band structure of semi-Heusler alloys was 
given using electronic structure calculations and tight-binding model
analysis,~\cite{deGroot83,Ogut,Galanakis,Nanda,Kulatov} and we briefly 
summarize the results. The key points, which determine the behavior 
of electrons near the Fermi level for the half-metallic property, are 
the interplay between the crystal structure, valence electron count, 
covalent bonding and large exchange splitting of the Mn-$d$ electrons.

\begin{figure}[t]
\rotatebox{270}{\includegraphics[height=\columnwidth]{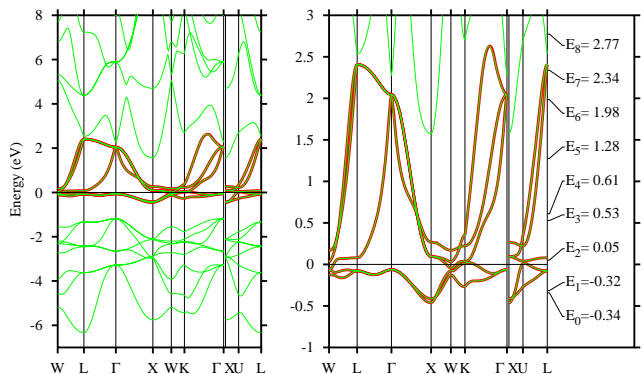}}
\caption{(Color online) 
The band structure of non-spin-polarized NiMnSb calculated with the 
full basis is given in thick (green) line. 
The thin (red) band have been calculated 
with a Mn-$d$ NMTO basis set. 
Fermi energy $E_F$ is set to be zero.
The high-symmetry points are
$W (\frac{1}{2}, 1, 0)$, $L (\frac{1}{2}, \frac{1}{2}, \frac{1}{2})$, 
$\Gamma (0, 0, 0)$, $X (0, 1, 0)$, $K (\frac{3}{4}, \frac{3}{4}, 0)$, 
in the $W$-$L$-$\Gamma$-$X$-$W$-$K$-$\Gamma$ line
and $X (0, 0, 1)$, $U (\frac{1}{4}, \frac{1}{4}, 1)$,
$L (\frac{1}{2}, \frac{1}{2}, \frac{1}{2})$, in the $X$-$U$-$L$ line.
Energy mesh used for downfolded calculation 
is given in the right of the band structure with a unit of eV.} 
\label{bnds_nsp}
\end{figure}

\begin{figure}[t]
\rotatebox{270}{\includegraphics[height=\columnwidth]{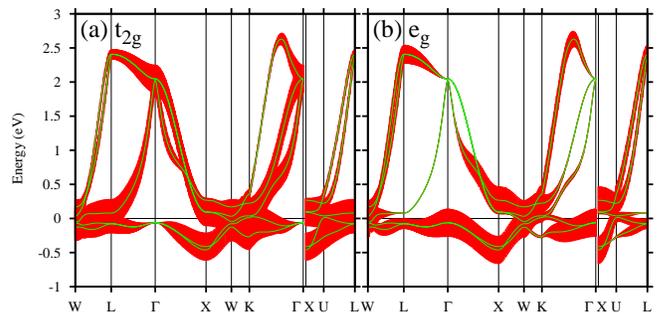}}
\caption{(Color online) 
The NMTO band structure calculated with a Mn-$d$ NMTO basis set. 
The bands have been decorated with (a) $t_{2g}$ and (b) $e_g$ characters.
About symmetry points and $E_F$, see Fig.~\ref{bnds_nsp}.
} 
\label{fatbnds_nsp}
\end{figure}

\begin{figure*}[t]
\includegraphics[width=\textwidth]{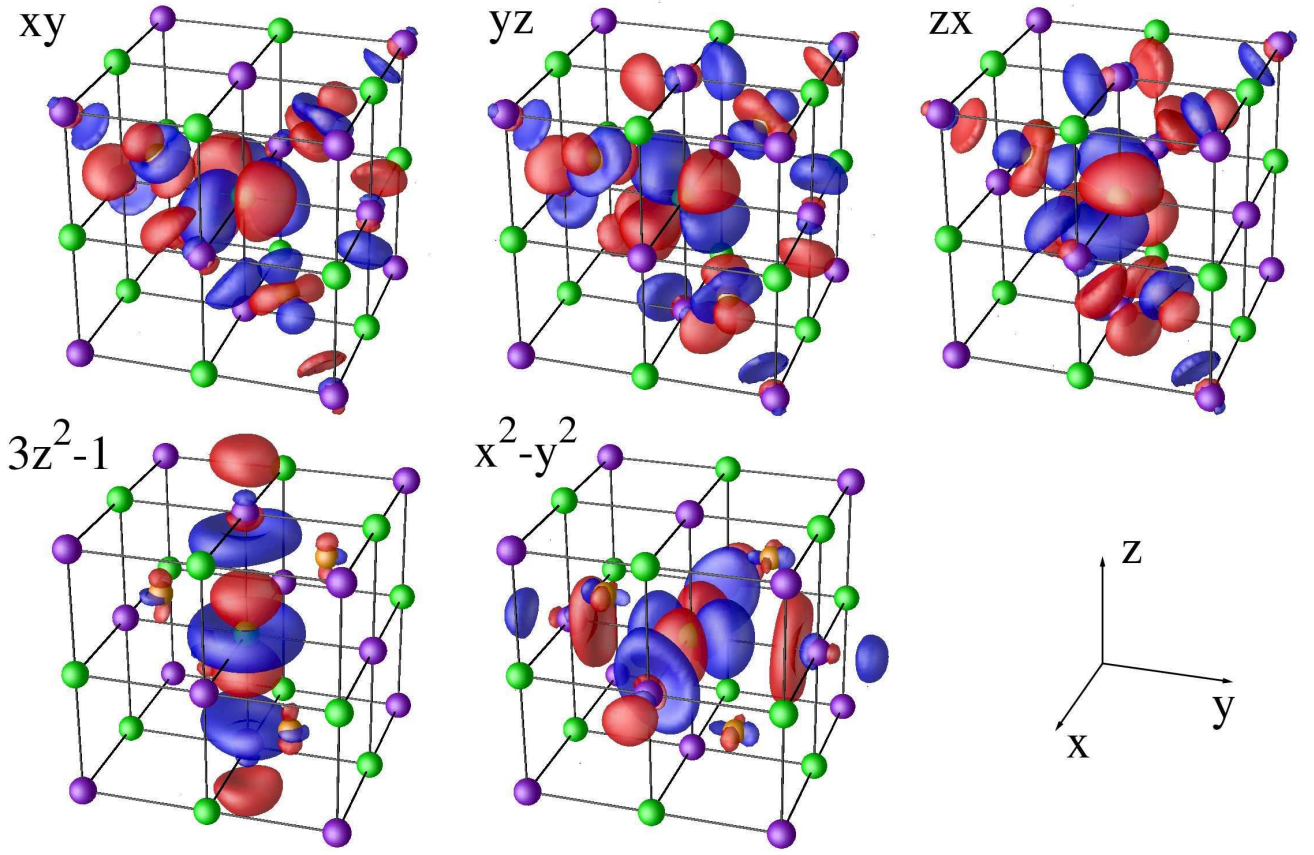}
\caption{(Color) NMTO Mn-$d$ Wannier orbitals of NiMnSb. 
Ni is orange, Mn is green, and Sb is purple.
Red (blue) indicates a positive (negative) sign.
Upper panel: $t_{2g}$ orbitals; $d_{xy}$ (left), $d_{yz}$ (middle), 
$d_{zx}$ (right). 
The triply degenerate $t_{2g}$ orbitals 
can be obtained by the permutation of axes.
Lower panel: $e_g$ orbitals; $d_{3z^2-1}$ (left), $d_{x^2-y^2}$ (middle).
These $e_g$ orbitals are doubly degenerated.} 
\label{Mndorbit}
\end{figure*}

\subsubsection{Chemical bonding and the {\rm Mn}-$d$ Wannier orbitals}

In the nonmagnetic phase the $C1_b$ compounds have a $d$-$d$ gap resulting 
from covalent hybridization of the higher-valent transition metal (Ni) 
with the lower-valent transition metal (Mn). For the minority spin gap 
opening, not only the Mn-$d$-Sb-$p$ interactions, but also 
Mn-$d$-Ni-$d$
interactions have to be taken into account. Moreover the loss of inversion 
symmetry produced by $C1_b$ structure (the symmetry lowering from $O_h$ 
in the $L2_1$ structure to $T_d$ in the $C1_b$ structure at Mn site) 
is an essential additional ingredient. 
All the above 
interactions combined with the $T_d$ symmetry lead to a nonzero anticrossing 
of bands and to the gap opening. The existence of $sp$-valent Sb 
is crucial to provide stability to this compound.

Based on the understanding of the bonding in the NiMnSb, we propose a 
downfolding scheme in which all orbitals of all atoms except Mn-$d$
is downfolded. 

In Fig. \ref{bnds_nsp} the non-spin-polarized band structure of NiMnSb 
is calculated with the full basis set (thick green line). 
Mn $d$, Ni $d$, and Sb $p$ states are lying between $-0.5$--$2.5$ eV, 
$-3$--$-1$ eV, and $-6.5$--$-3$ eV, respectively.
Sb $s$ states are sitting around $-12$--$-10$ eV, 
that is not shown in Fig. \ref{bnds_nsp}.
There is an excellent agreement 
with the previous calculations.~\cite{deGroot83}
The thin red band has been calculated 
with the downfolded basis set which includes only  Mn-$d$ orbitals. The energy 
mesh used in the downfolded calculation is given to the right of the band structure. 
The two sets of bands are 
identical. 
In Fig. \ref{fatbnds_nsp} the bandstructures of NiMnSb with the orbital
character projected on to NMTO $t_{2g}$ and $e_g$ Wannier orbitals are shown.
The fatness associated with each band is proportional to the character of the
orbital.
The strong hybridization between $t_{2g}$ and $e_g$ states is clearly seen.
NMTO Mn-$d$ Wannier orbitals are shown in Fig.~\ref{Mndorbit}. 
The triply degenerate manganese $t_{2g}$ orbitals are very complicated due to the 
hybridization with Ni-$d$ and Sb-$p$ states. The $d_{xy}$ orbital at Mn 
site is deformed by antibonding with the Ni-$d$ state directed tetrahedrally to 
[$11\bar{1}$], [$\bar{1}\bar{1}\bar{1}$], [$1\bar{1}1$], and [$\bar{1}11$].
The same Ni-$d$ orbitals couple with Sb-$p$ states. The direct 
Mn-$d_{xy}$-Sb-$p$ $\pi$ coupling is not seen since the distance is 
$d$(Mn-Sb):$d$(Ni-Sb)=1:$\frac{\sqrt{3}}{2}$. Therefore the Ni-$d$-Sb-$p$ 
interactions are more favorable. The dispersion of the Mn $t_{2g}$ bands 
is mainly due to hopping via the tails of Sb-$p$ and Ni-$d$ orbitals. 
On the other hand, the second nearest neighbor (NN) $d$-$d$ hopping of $t_{2g}$ 
orbital is small. The $e_g$ orbitals at Mn site are much easier to understand: 
they point towards Sb atoms, and a strong $pd\sigma$ coupling between 
Sb-$p$ and Mn-$e_g$ states is seen.
This induces large second NN $d$-$d$ hoppings.

\subsubsection{Effective hopping matrix elements in the basis set of {\rm Mn}-$d$ Wannier orbitals}

In the many body picture the Mn $t_{2g}$ and $e_{g}$ constitutes 
the {\it active orbitals} which are responsible for the low energy physics, 
having fluctuation in occupation and spins. The effective hopping Hamiltonian 
matrix elements built up with these active orbitals are as follows.

NMTO basis set:
\begin{equation}
\left| \chi^{\perp} \right\rangle =
\left\{ |xy \rangle, |yz\rangle, |zx\rangle, |3z^2{\rm -}1\rangle, 
|x^2{\rm -}y^2\rangle  \right\}.
\label{Heff_NiMnSb_basis}
\end{equation}

The on-site term:
\begin{equation}
t_{m',m}^{000} = \left(
\begin{array}{rrrrr}
 360 &    0 &    0 &    0 &    0 \\
   0 &  360 &    0 &    0 &    0 \\
   0 &    0 &  360 &    0 &    0 \\
   0 &    0 &    0 &  434 &    0 \\
   0 &    0 &    0 &    0 &  434
\end{array}
\right).
\label{Heff_NiMnSb_0nn}
\end{equation}

The first nearest neighbor:
\begin{equation}
t_{m',m}^{0\frac{1}{2}\frac{1}{2}} = \left(
\begin{array}{rrrrr}
 129 &    2 &   23 & -100 &   12 \\
  -2 &   51 &   -2 &   75 & -130 \\
  23 &    2 &  129 &   40 &   92 \\
 100 &   75 &  -40 &    9 &  -37 \\
 -12 & -130 &  -92 &  -37 &   51
\end{array}
\right). 
\label{Heff_NiMnSb_1nn}
\end{equation}

The second nearest neighbor:
\begin{equation}
t_{m',m}^{001} = \left(
\begin{array}{rrrrr}
 -16 &    0 &    0 &  -43 &    0 \\
   0 &   -4 &    0 &    0 &    0 \\
   0 &    0 &   -4 &    0 &    0 \\
  43 &    0 &    0 & -229 &    0 \\
   0 &    0 &    0 &    0 &   -4
\end{array}
\right),
\label{Heff_NiMnSb_2nn}
\end{equation}
where the unit is meV, $E_F=0$, and hopping integrals up to the second 
NN are shown. The on-site term $t_{m',m}^{000}$ is diagonal, $t_{2g}$ 
and $e_g$ orbitals are triply and doubly degenerated, respectively. The 
crystal-field splitting between $t_{2g}$ and $e_g$ orbitals is $\sim 74$ 
meV. There are 12 first NN and 6 second NN hoppings. Only one hopping 
integral at each NN is shown, but all the hopping integrals can be derived 
from proper unitary transformation due to the crystal symmetry. For details, 
see Ref.~[\onlinecite{PavariniNJP}]. The hopping between $t_{2g}$ and $e_g$ 
orbitals is strongly influenced by the presence of tails belonging to the 
Sb-$p$ and Ni-$d$ orbitals. Due to these tails large values of $t_{2g}$ 
to $e_g$ hoppings are evidenced in Eqs. 
(\ref{Heff_NiMnSb_1nn}) and (\ref{Heff_NiMnSb_2nn}). 
We mention that these 
hoppings should be small for the case of ``pure'' $t_{2g}$ and $e_g$ orbitals. 
In addition, due to the presence of Sb-$p$ and Ni-$d$ tails  
$t_{2g}$ and $e_g$ orbitals are not divided clearly as seen in Fig.~\ref{fatbnds_nsp}. 
Therefore they should be treated equally. 
Hoppings further than third NN are small; for instance, 
the largest values are 66 (third) and 56 meV (fourth), a small number of 
matrix elements of the hoppings are less than $\sim$30 meV, and the others are 
almost zero in the third and fourth NN hoppings. Further NN hoppings are 
negligible. 

In this procedure, we obtained the nonzero hopping matrix elements between 
the all Mn-$d$ orbitals, in the downfolded representation. The numbers of 
independent parameters of the effective model are equal to the number
of matrix elements from Eqs. (\ref{Heff_NiMnSb_0nn})--(\ref{Heff_NiMnSb_2nn}). 

We now consider the spin-polarized case. Systems with 18 valence
electrons per unit cell are semiconductors, but when they contain more than 18
electrons (22 electrons in NiMnSb, that is, Ni 3$d^8$4$s^2$, Mn 3$d^5$4$s^2$,
and Sb 5$s^2$5$p^3$), antibonding states are populated.
Therefore, the nonmagnetic phase becomes unstable and the magnetic state
can be stabilized. The large exchange splitting of the Mn atom 
(producing a magnetic moment of around $\sim$3.7 $\mu_B$) is crucial to induce 
a half-metallic property. As we saw before, for the nonmagnetic case, the $t_{2g}$ 
and  $e_{g}$ states of Ni are situated around 2 eV below $E_F$. 
In the spin-polarized 
calculation their position is slightly changed, therefore the exchange splitting
on Ni is not large. The actual magnetic moment calculation gives a value
around $0.3\mu_B$.

\begin{figure}[t]
\rotatebox{270}{\includegraphics[height=\columnwidth]{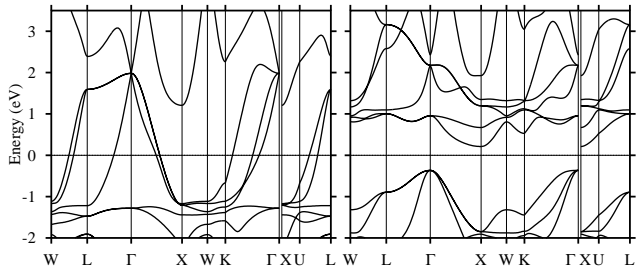}}
\caption{Full basis set spin-polarized (ferromagnetic) bands for NiMnSb;
majority spin (left) and minority spin (right).
About symmetry points and $E_F$, see Fig.~\ref{bnds_nsp}.
The similarity of the spin-polarized majority   
with the non-spin-polarized bands is evidenced.}
\label{bnds_sp}
\end{figure}

The non-spin-polarized result has a striking resemblance to the majority
spin-polarized calculations, presented in Fig. \ref{bnds_sp}. Kulatov
{\it et al.} explained half-metallicity of NiMnSb and CrO$_2$ by the
extended Stoner factor calculations in the rigid-band approximation:~\cite{Kulatov}
the minority spin-band gap opens due to the exchange splitting, which shifts minority 
bands, so they become empty. According to our results, the NMTO antibonding orbitals 
should be a good description for the empty Mn-$d$ states in the minority channel.

All the above results suggest that a minimal basis set which captures the 
essential physics in NiMnSb can be constructed using Mn $t_{2g}$ and $e_{g}$ 
states. Therefore, the non-spin-polarized result with Mn-$d$ orbitals can be a good 
starting point for the many-body calculations.

\begin{figure}[t]
\includegraphics[width=\columnwidth]{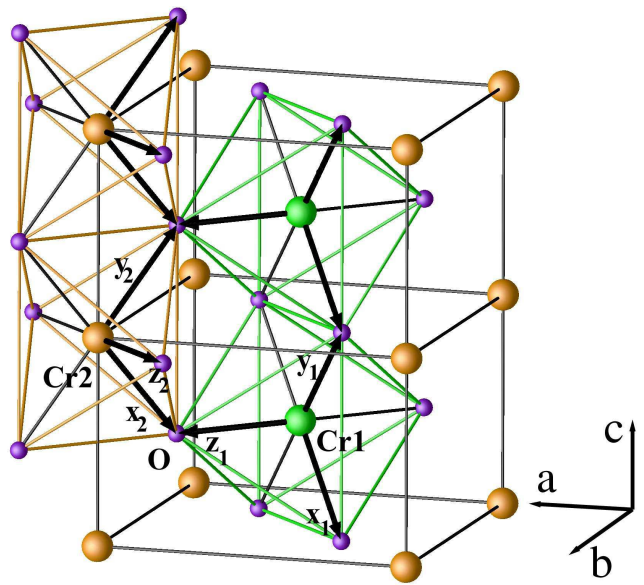}
\caption{(Color online) 
CrO$_2$ (rutile) structure. 
Cr1 (green) and Cr2 (orange) are located at (0, 0, 0) and 
$(\frac{1}{2}, \frac{1}{2}, \frac{1}{2})$. 
Cr atoms are octahedrally coordinated by oxygen atoms (purple).
The local coordinate system is used for each Cr atom;
$\hat{{\bf x}}_1=-\frac{1}{2}\hat{{\bf a}}
                 +\frac{1}{2}\hat{{\bf b}}
                 -\frac{1}{\sqrt{2}}\hat{{\bf c}}$,
$\hat{{\bf y}}_1=-\frac{1}{2}\hat{{\bf a}}
                 +\frac{1}{2}\hat{{\bf b}}
                 +\frac{1}{\sqrt{2}}\hat{{\bf c}}$,
$\hat{{\bf z}}_1= \frac{1}{\sqrt{2}}\hat{{\bf a}}
                 +\frac{1}{\sqrt{2}}\hat{{\bf b}}$, and
$\hat{{\bf x}}_2=-\frac{1}{2}\hat{{\bf a}}
                 -\frac{1}{2}\hat{{\bf b}}
                 -\frac{1}{\sqrt{2}}\hat{{\bf c}}$,
$\hat{{\bf y}}_2=-\frac{1}{2}\hat{{\bf a}}
                 -\frac{1}{2}\hat{{\bf b}}
                 +\frac{1}{\sqrt{2}}\hat{{\bf c}}$,
$\hat{{\bf z}}_2=-\frac{1}{\sqrt{2}}\hat{{\bf a}}
                 +\frac{1}{\sqrt{2}}\hat{{\bf b}}$.
$\hat{{\bf x}}_{1,2}$ and $\hat{{\bf y}}_{1,2}$ are approximately point to O atom,
and $\hat{{\bf z}}_{1,2}$ are exactly point to O atom.
The local axis is transformed into each other by a rotation of 90$^\circ$
around the crystal $c$ axis.} 
\label{CrO2struc}
\end{figure}

\subsection{Downfolding onto the Cr-$d$ manifolds in CrO$_2$\label{downcro2}}

Chromium dioxide CrO$_2$ has a rutile (tetragonal) structure with $a$=$4.421$ \AA, 
$c$=$2.916$ \AA ~($c/a$=$0.65958$), and internal parameter $u$=$0.3053$.~\cite{CrO2str}
The Cr atoms form a 
body-center tetragonal lattice and are surrounded by a slightly distorted octahedron 
of oxygen atoms. The space group of this compound is nonsymmorphic 
($P4_2/mnm=D_{4h}^{14}$). Cr$^{4+}$ has a close shell Ar core and two 
additional $3d$ electrons. The Cr ions are in the center of the CrO$_6$ 
octahedra so the $3d$ orbitals are split into a $t_{2g}$ triplet and an excited $e_{g}$ 
doublet. With only two $3d$ electrons the $e_{g}$ states are irrelevant and only the 
$t_{2g}$ orbitals need to be considered. The tetragonal symmetry distorts the octahedra, 
which lifts the degeneracy of the $t_{2g}$ orbitals into a $d_{xy}$ ground state and 
$d_{yz+zx}$ and $d_{yz-zx}$ excited states,~\cite{Lewis97,Korotin98} where {\it a local
coordinate system} is used for every octahedron (see Fig. \ref{CrO2struc}).
A double exchange mechanism for the two electrons per Cr site was 
proposed~\cite{Schlottmann03} 
in which due to the strong Hund's rule corroborated with the 
distortion of CrO$_6$ octahedra leads to the localization of the one electron 
into the $d_{xy}$ orbital, while the electrons in the $d_{yz}$ and $d_{xz}$ are itinerant.

Measurements of the magnetic susceptibility in the paramagnetic phase show a 
Curie-Weiss-like behavior indicating the presence of local moments,~\cite{Chamberland77} 
suggesting a mechanism of ferromagnetism beyond the standard band or Stoner-like model.

Several recent experiments including photoemission,~\cite{Tsujioka97} soft x-ray
absorption,~\cite{Stagarescu00} resistivity,~\cite{Suzuki98} and optics~\cite{Singley99}
suggest that electron correlations are essential to the underlying physical picture
in CrO$_2$. 
Schwarz~\cite{Schwarz86} 
first predicted the half-metallic band structure with a spin moment of $2 \mu_B$ per 
formula unit for CrO$_2$. Later on Lewis~\cite{Lewis97} used the plane-wave potential 
method and investigated the energy bands and the transport properties, characterizing 
CrO$_2$ as a ``bad metal,'' 
terminology applied to high temperature superconductors, 
or to the other transition metal oxides, even ferromagnets such as SrRuO$_3$. A 
decade later the LSDA+$U$ calculation~\cite{Korotin98} explained the conductivity in the 
presence of large on-site Coulomb interactions, and concluded that CrO$_2$ is a 
negative charge transfer gap material which leads to self-doping. Contrary to the 
on-site strong correlation description, transport and optical properties obtained within 
the FLAPW method (LSDA and GGA),~\cite{Mazin99} suggest that the electron-magnon scattering 
is  responsible for the renormalization of the one-electron bands. In addition, more recent 
model calculations proposed even orbital correlations.~\cite{Hartmann03}

\begin{figure}[t]
\rotatebox{270}{\includegraphics[height=\columnwidth]{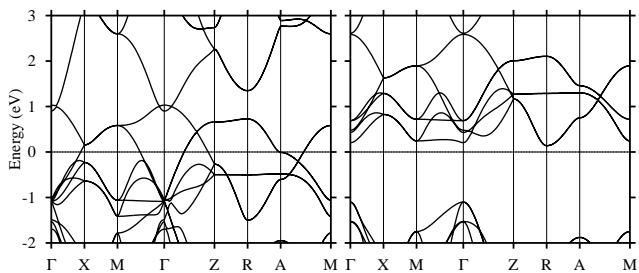}}
\caption{Full basis set spin-polarized (ferromagnetic) bands for CrO$_2$;
majority spin (left) and minority spin (right).
$E_F$ is set to be zero.
The high-symmetry points are
$\Gamma (0, 0, 0)$, $X (0, \frac{1}{2}, 0)$, $M (\frac{1}{2}, \frac{1}{2}, 0)$,
$Z (0, 0, \frac{1}{2})$, $R (0, \frac{1}{2}, \frac{1}{2})$, 
$A (\frac{1}{2}, \frac{1}{2}, \frac{1}{2})$. 
}
\label{bnds_sp_cro2}
\end{figure}

\begin{figure}[b]
\rotatebox{270}{\includegraphics[height=\columnwidth]{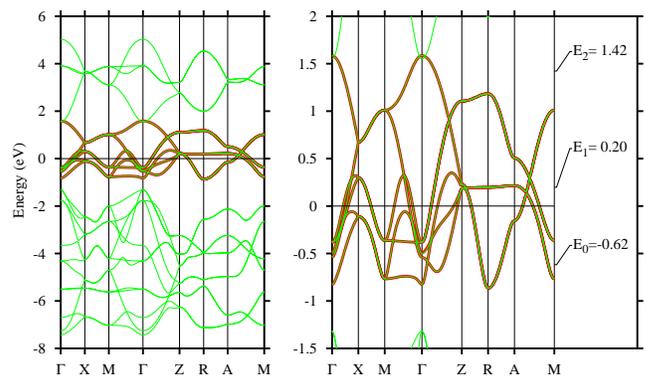}}
\caption{(Color online) 
The band structure of non-spin-polarized CrO$_2$ calculated with the 
full basis is given in thick (green) line. 
The thin (red) band have been calculated 
with a Cr-$t_{2g}$ NMTO basis set. 
Energy mesh used for downfolded calculation 
is given in the right of the band structure with a unit of eV.
About symmetry points and $E_F$, see Fig.~\ref{bnds_sp_cro2}.}
\label{bnds_nsp_cro2}
\end{figure}

\begin{figure*}[t]
\includegraphics[width=\textwidth]{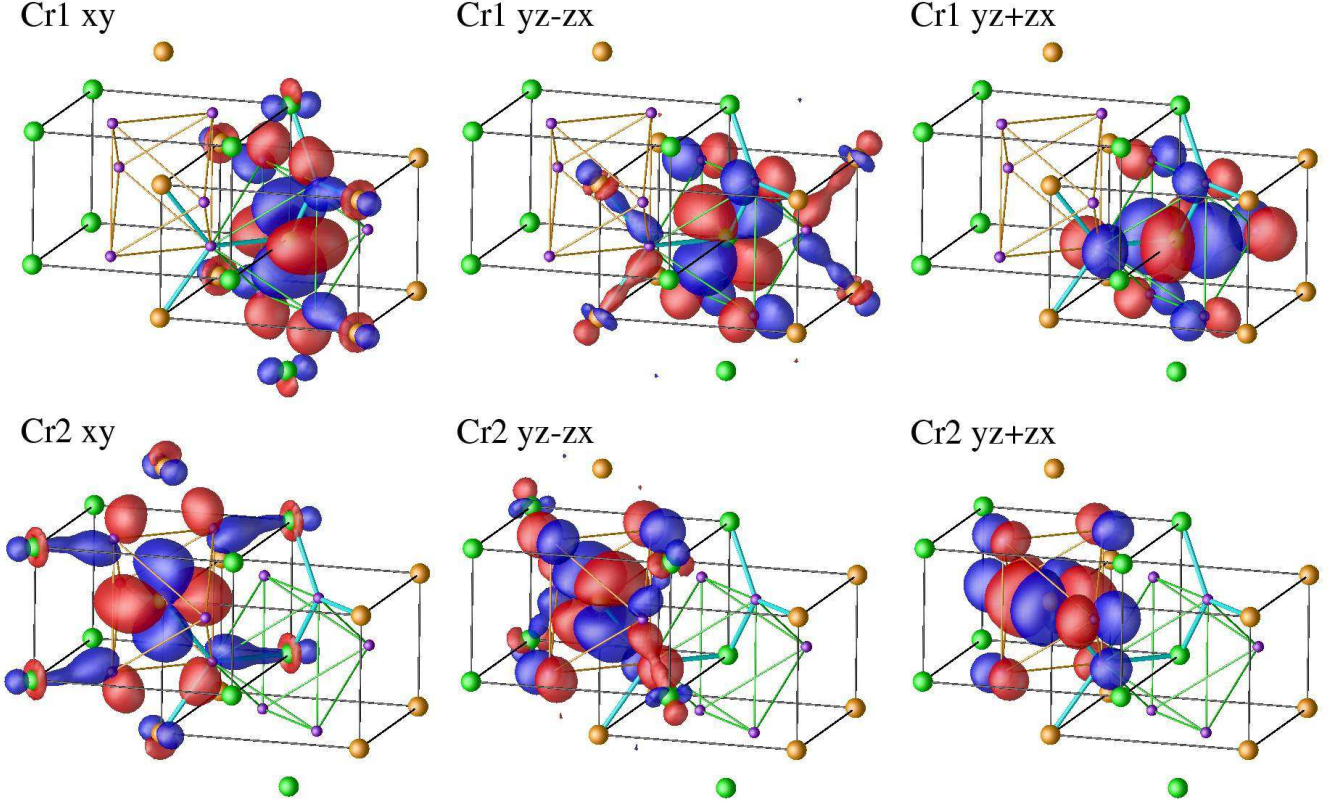}
\caption{(Color) 
NMTO Cr-$t_{2g}$ Wannier orbitals of CrO$_2$.
Cr1 is orange, Cr2 is green, and O is purple.
The local coordinate systems are used for each atom.
Red (blue) indicates a positive (negative) sign.
Upper panel: Cr1 $t_{2g}$ orbitals.
Lower panel: Cr2 $t_{2g}$ orbitals.
$d_{xy}$ (left). This orbital is singly degenerated.
$d_{yz-zx}$ (middle) and 
$d_{yz+zx}$ (right). These two orbitals are nearly degenerated.
} 
\label{Crt2gorbit}
\end{figure*}

\subsubsection{Chemical bonding and the {\rm Cr}-$t_{2g}$ Wannier orbitals}
Chemical bonding in rutile-type compounds including CrO$_2$ was analyzed by 
Sorantin and Schwarz.~\cite{Sorantin92} Let us summarize their results: one can see
that around the Fermi level, the bands are primarily chromium $3d$ states of $t_{2g}$ manifold, 
with $e_g$ bands situated higher in energy due to the crystal-field splitting. 
In the spin 
polarized case, the exchange splitting shifts the minority spin-$d$ bands above the 
Fermi level, as seen in Fig.~\ref{bnds_sp_cro2}. For the majority $t_{2g}$ bands the 
Fermi level is lying in a pseudogap. Oxygen $p$-chromium $d$, hybridization creates 
both bonding and antibonding hybrid orbitals, with the bonding orbital appearing 
in the occupied part and the antibonding hybrid orbital remaining in the Cr $t_{2g}$ 
manifold. Half of the $d_{yz}$ and $d_{zx}$ components of $t_{2g}$ are pushed upward, by 
antibonding, that explains the dominance of $d_{xy}$ character in the spin density. 
The non-magnetic density of states shows a sharp peak at the Fermi level, which 
signals the magnetic instability according to the usual Stoner argument. 

By the procedure of NMTOs we can easily gain a more complete picture about the chemical 
bonding. 
In Fig. \ref{bnds_nsp_cro2} the non-spin-polarized band structure of CrO$_2$ is 
calculated with the full basis set (thick green line). 
Cr $e_g$, Cr $t_{2g}$, and O $p$ states are lying between $1.5$--$5$ eV, 
$-1$--$1.5$ eV, and $-7.5$--$-1.5$ eV, respectively.
There is an excellent agreement with
the previous calculations.~\cite{Korotin98,Schwarz86,Mazin99,Sorantin92}
The thin red bands have been
calculated with the downfolded basis set which includes only  Cr-$d$ orbitals. 
The energy mesh used in the downfolded calculation is given to the right of the band structure.
NMTO Cr-$t_{2g}$ Wannier orbitals at Cr1 (0, 0, 0) and 
Cr2 ($\frac{1}{2}$, $\frac{1}{2}$, $\frac{1}{2}$) are shown in Fig.~\ref{Crt2gorbit}. 
All $t_{2g}$ orbitals form antibonding $pd\pi$ coupling with O-2$p$ states.
These antibonding NMTO Wannier orbitals capture the essentials to describe the 
half-metallicity of CrO$_2$. 

Note that the local coordinate system is used.~\cite{note}
For instance, the $z$ axis at Cr1 site 
points along the [$110$], while the $z$ axis at Cr2 sites points towards the 
[$\bar{1}10$] direction.
In the following we describe the essential features of the Cr-$d$ and 
O-2$p$ orbital couplings in the $t_{2g}$ orbitals on Cr1 site: 

(i) Cr$_3$O {\em cluster.}
The oxygen atoms yield a Cr$_3$O cluster with three surrounding Cr atoms.
The Cr1 atom has six Cr$_3$O clusters with two different types; for instance,
one Cr$_3$O cluster is formed by 
Cr1 at (0,0,0) and (0,0,1), Cr2 at ($-\frac{1}{2}$,$ \frac{1}{2}$,$\frac{1}{2}$), and
O at ($u-\frac{1}{2}$,$\frac{1}{2}-u$,$\frac{1}{2}$) in the (110) plane. 
Another type of Cr$_3$O cluster is formed by 
Cr1 at (0,0,0), Cr2 at ($\frac{1}{2}$,$ \frac{1}{2}$,$\pm\frac{1}{2}$), and
O at ($u$,$u$,$0$) in the ($\bar{1}10$) plane. 
The O is sitting in the center of 
a triangle formed by the three coplanar 
Cr nearest atoms. The Cr$_3$O cluster is indicated by thick cyan lines
in Fig.~\ref{Crt2gorbit}. 
In the Cr$_3$O unit, one can see one antibonding $pd\pi$ coupling between 
Cr-$t_{2g}$ and O-$p$ states and two bonding $pd\sigma$ couplings 
between Cr-$e_g$ and O-$p$ states. On one hand, the Cr-$e_g$ 
and Cr-$t_{2g}$ orbitals lying within the Cr$_3$O plane couple 
to the in-plane O-$p$ orbital. On the other hand, the out-of-plane 
O $p$ orbital component, perpendicular to the Cr$_3$O plane, cannot couple
to the Cr-$e_g$ state due to orthogonality. 

(ii) {\em $d_{xy}$ orbitals.}
The Cr1 atom experiences a bonding coupling between its $d_{xy}$ orbital 
and the $e_g$ ($d_{3z^2-1}$) orbital located on the other six nearest
Cr atoms on the (110) plane via Cr$_3$O cluster.
The bonding is realized via
the tails of O-2$p$ ($p_x$- and $p_y$-like) orbitals. 
$e_g$ tails belonging to the first NN Cr1 (0,0,$\pm$1) and second NN
Cr2 ($-\frac{1}{2}$,$ \frac{1}{2}$,$\pm\frac{1}{2}$),
    ($ \frac{1}{2}$,$-\frac{1}{2}$,$\pm\frac{1}{2}$) atoms
are visible in the $(110)$ plane, where the $d_{xy}$ orbital is situated.  
Further, the $d_{3z^2-1}$ state belonging to the Cr1 (0,0,1) atom forms a
$pd\sigma$-type bonding with both O $p_x$ orbital located at 
($u-\frac{1}{2}$,$\frac{1}{2}-u$,$\frac{1}{2}$) and with O $p_y$ situated 
at ($\frac{1}{2}-u$,$u-\frac{1}{2}$,$\frac{1}{2}$). 
The $d_{xy}$ orbital, on the contrary, forms an antibonding coupling with these 
O $p_x$ and O $p_y$ states.
Cr2 atom situated at ($-\frac{1}{2}$,$ \frac{1}{2}$,$\frac{1}{2}$) involves 
its $d_{3z^2-1}$ orbital into a $pd\sigma$ bonding with O $p_x$ orbital 
situated at ($u-\frac{1}{2}$,$\frac{1}{2}-u$,$\frac{1}{2}$). 

(iii) {\em $d_{yz-zx}$ orbitals.}
The O atoms situated at ($u$,$u$,0) and ($-u$,$-u$,0) intermediate a
$pd\sigma$ bonding, via their $p_{x-y}$ orbital, between the $d_{yz-zx}$ 
orbital at Cr1 (0,0,0) and
the $e_g$ orbitals at Cr2 ($\frac{1}{2}$,$ \frac{1}{2}$,$\pm\frac{1}{2}$) and 
($-\frac{1}{2}$,$-\frac{1}{2}$,$\pm\frac{1}{2}$) atoms. 
This bonding orbital is situated in the $(1\bar{1}0)$ plane similarly to the 
$d_{yz-zx}$ orbital. However, its bonding strength seems to be weaker than the 
ones of the $d_{xy}$ orbitals, due to a larger distance between O and   
Cr2 atoms. 
The  O $p_z$ orbitals belonging to the atoms situated at 
($\frac{1}{2}-u$,$u-\frac{1}{2}$,$\pm\frac{1}{2}$) and
($u-\frac{1}{2}$,$\frac{1}{2}-u$,$\pm\frac{1}{2}$),
form a $pd\pi$ coupling with $d_{yz-zx}$ orbital. Therefore 
the O $p_{z}$ orbital perpendicular to the plane of the 
Cr$_3$O cluster does not overlap with the $e_g$ tails of 
Cr1 (0,0,$\pm 1$) atoms sitting along the [001] direction. 

(iv) {\em $d_{yz+zx}$ orbitals.}
The $e_g$ tails can not contribute at all to the $d_{yz+zx}$ orbital. 
Because O $p_z$ orbitals at 
($\frac{1}{2}-u$,$u-\frac{1}{2}$,$\pm\frac{1}{2}$),
($u-\frac{1}{2}$,$\frac{1}{2}-u$,$\pm\frac{1}{2}$) and O $p_{x+y}$ orbitals at
($u$,$u$,0), ($-u$,$-u$,0)
are situated perpendicular to Cr$_3$O clusters,
the coupling between O $p$ ($p_{z}$ and $p_{x+y}$) states and $e_g$ states
at surrounding Cr atoms is not allowed due to the orthogonality.

We discussed in the above points (i)--(iv), the direct or mediated interactions 
between Cr-$d$ and O-$p$ or between the Cr $t_{2g}$ and $e_g$ 
states. As it is already known,~\cite{Korotin98,Schwarz86,Mazin99,Sorantin92}
there is a significant difference between the $t_{2g}$ and $e_g$ 
orbitals, however the above analysis in the framework of NMTO technique shows 
that their interplay constitutes an important ingredient not only 
for the crystal-field splitting of $t_{2g}$ states, but also for the general  
bonding in the rutile structure. The $t_{2g}$ orbitals form the basis set in 
which the effective hopping Hamiltonian matrix elements are evaluated. These
results are presented below.

\subsubsection{Effective hopping matrix elements in the {\rm Cr}-$t_{2g}$ Wannier orbitals}

The hopping integrals $t$ with $t_{2g}$ Wannier representation up to the 
second NN are as follows.

NMTO basis set:
\begin{equation}
\left| \chi^{\perp} \right\rangle =
\left\{ |xy \rangle, |yz \!-\! zx\rangle, |yz \!+\! zx\rangle\right\}.
\label{Heff_CrO2_basis}
\end{equation}

The on-site term:
\begin{equation}
t_{m',m}^{000}
{\rm =} \left(
\begin{array}{rrr}
 104 &    0 &    0 \\
   0 &  323 &    0 \\
   0 &    0 &  355 
\end{array}
\right).
\label{Heff_CrO2_0nn}
\end{equation}

The first nearest neighbor:
\begin{equation}
t_{m',m}^{001 ({\rm Cr1}\rightarrow{\rm Cr1})} 
{\rm =} \left(
\begin{array}{rrr}
-119 &    0 &    0 \\
   0 & -177 &    0 \\
   0 &    0 &  196 
\end{array}
\right).
\label{Heff_CrO2_1nn}
\end{equation}

The second nearest neighbor:
\begin{equation}
t_{m',m}^{\frac{1}{2}\frac{1}{2}\frac{1}{2} ({\rm Cr1}\rightarrow{\rm Cr2})} 
{\rm =} \left(
\begin{array}{rrr}
  -4 &    0 &    0 \\
  32 &    0 &    0 \\
   0 & -204 &  142 
\end{array}
\right),
\label{Heff_CrO2_2nn}
\end{equation}
where the unit is meV and {\it a local coordinate system} for each atom Cr1 
and Cr2 in the unit cell is used as seen in Fig.~\ref{CrO2struc}.
The on-site term $t_{m',m}^{000}$ and the first NN hoppings
are diagonal in a $\{ d_{xy},d_{yz-zx},d_{yz+zx} \}$
representation. The splitting between these orbitals is due partly to the orthorhombic 
distortion of CrO$_6$ octahedra and partly to the bonding with the $e_g$ states 
of nearest Cr atoms in rutile structure. As a consequence mainly two crystal-field
levels are formed, a single $d_{xy}$ and two nearly degenerate $d_{yz\pm zx}$ orbitals
situated with $\sim 235$ meV at higher energies, that can be derived from the difference
between on-site levels of $d_{xy}$ and $d_{yz\pm zx}$ orbitals 
in Eq. (\ref{Heff_CrO2_0nn}). 
A possible ingredient in determining
the position of the nearly degenerate $d_{yz\pm zx}$ orbitals with respect to the 
single $d_{xy}$ orbital is the bond length between Cr and O atoms. 
In the calculations presented in this paper a longer  bond length along $z$ direction
($d_{\perp}=1.91$ \AA) is used in comparison with the bond formed in the  $xy$ plane
having a value of ($d_{\parallel}=1.89$ \AA). In another experiment, however, the 
opposite situation is reported, $d_{\perp}=1.89$ \AA{} is smaller than 
$d_{\parallel}=1.91$ {} \AA.~\cite{CrO2str2} 
In both cases, however, the band structure and density of states are almost identical.
Moreover other oxides (TiO$_2$, VO$_2$ etc.) with rutile structure have a
similar band structure and density of states for $t_{2g}$ states.
This fact suggests that the bonding with $e_g$ state is essential for the crystal-field
splitting of $t_{2g}$ states.
Similar mechanism for the crystal-field splitting of $t_{2g}$ state
is realized in V$_2$O$_3$ (Ref.~\onlinecite{V2O3}) and NaCoO$_2$.~\cite{NaCoO2}

\begin{figure}[t]
\rotatebox{270}{\includegraphics[height=\columnwidth]{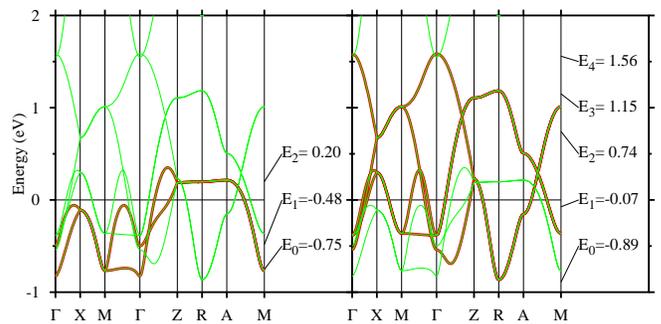}}
\caption{(Color online) 
NMTO downfolded $d_{xy}$ (left) and $d_{yz\!\pm\!zx}$ (right) 
bands for CrO$_2$.
Energy mesh used for each calculation 
is given in the right of the band structure with a unit of eV.
For symmetry points and $E_F$, see Fig.~\ref{bnds_sp_cro2}.}
\label{bnds_nmto_cro2}
\end{figure}

There are two first NN between Cr1 (Cr2) and eight second 
NN hoppings between Cr1 and Cr2. Only one hopping integral at 
each NN is shown here. The hopping integrals are strongly affected by Wannier 
function's tails which are mainly due to downfolded O 2$p$ and Cr $e_g$ 
orbitals. $d_{xy}$ orbital has a small second nearest hopping integrals
$t_{xy,xy}^{\frac{1}{2}\frac{1}{2}\frac{1}{2}}$=$-4$ meV although O $p$ tails are present.
This means that $d_{xy}$ orbital does not have the second NN hopping path, indicating 
a very narrow $d_{xy}$ band. On contrary, large hopping integrals of the second NN 
are shown in $d_{yz\pm zx}$ block. $d_{yz+zx}$ orbitals can hop to $d_{yz+zx}$ of 
all direction ($\left| t_{yz+zx,yz+zx}^{{\rm 2nd \mbox{ } NN}} \right| $=$142$ meV),
but $d_{yz-zx}$ can not to $d_{yz-zx}$ ($t_{yz-zx,yz-zx}^{{\rm 2nd \mbox{ } NN}}$=$0$ 
meV exactly, as a consequence of orthogonality.) The hopping between $d_{yz-zx}$ and $d_{yz+zx}$ 
($|t|$=204 meV) is allowed for a particular direction, that is, the second NN hopping 
from Cr1 $d_{yz+zx}$ to Cr2 $d_{yz-zx}$ sitting in ($1\bar{1}0$) plane or from Cr1 
$d_{yz-zx}$ to Cr2 $d_{yz+zx}$ sitting in ($110$) plane.
The large values of second NN hopping in $d_{yz\pm zx}$ block 
are caused by O $p$ tails which are not orthogonal to each other,
and induce both the itinerant property and the wide band of $d_{yz\pm zx}$ states.
$d_{yz\pm zx}$ orbitals have the second NN hopping path via O $p$ tails.  
Moreover the large off-diagonal hopping integral of second NN and 
the small crystal-field splitting between $d_{yz\pm zx}$
indicate that $d_{yz\pm zx}$ states should be treated together.
The Cr1 $d_{yz-zx}$ orbital is weakly coupled with Cr2 $d_{xy}$ ($|t|$=32 meV) 
along particular directions, within the ($1\bar{1}0$) plane.  
Hoppings towards the third NN are small; for instance, the largest 
values are 45 meV (third) and 30 meV (fourth and fifth), and the others are 
almost zero. Further away NN hoppings are much smaller.

Exceptions are, however, the $e_g$ tails of Wannier function which gives a 
few large hopping integrals, for instance, $t_{xy,xy}^{002 ({\rm Cr1}\rightarrow{\rm Cr1})}=-82$ meV, 
$t_{xy,xy}^{1\bar{1}0 ({\rm Cr1}\rightarrow{\rm Cr1})}=-95$ meV, and
$t_{yz-zx,yz-zx}^{11\frac{1}{2} ({\rm Cr1}\rightarrow{\rm Cr1})}=-50$ meV.
These large hoppings can be seen in Fig.~\ref{Crt2gorbit}.

\subsubsection{Effective hopping matrix elements in the independent $d_{xy}$ 
and $d_{yz\pm zx}$ Wannier orbitals}

As seen in Eqs. (\ref{Heff_CrO2_0nn})--(\ref{Heff_CrO2_2nn}), 
the coupling between $d_{xy}$ and $d_{yz\pm zx}$ 
seems to be weak. The NMTO method may be able to pick up such bands independently. 
The $d_{xy}$ and 
$d_{yz\pm zx}$ bands are shown in Fig.~\ref{bnds_nmto_cro2}. The Wannier orbitals are 
more extended due to additional downfolding. This affects the hopping integrals as well.
The hopping integrals $\tilde{t}$ with $d_{xy}$ and $d_{yz\pm zx}$ Wannier 
representation up to the second NN are as follows.

NMTO basis set:
\begin{equation}
\left| \chi^{\perp} \right\rangle =
\bigl\{ \{|xy \rangle\}, \{|yz \!-\! zx\rangle, |yz \!+\! zx\rangle \} \bigr\}.
\label{Heff_CrO2_sep_basis}
\end{equation}

The on-site term:
\begin{equation}
\tilde{t}_{m',m}^{000} 
{\rm =} \left(
\begin{array}{rrr}
 -24 &      &      \\
     &  312 &    0 \\
     &    0 &  373 
\end{array}
\right).
\label{Heff_CrO2_sep_0nn}
\end{equation}

The first nearest neighbor:
\begin{equation}
\tilde{t}_{m',m}^{001 ({\rm Cr1}\rightarrow{\rm Cr1})} 
{\rm =} \left(
\begin{array}{rrr}
 -99 &      &      \\
     & -187 &    0 \\
     &    0 &  197 
\end{array}
\right).
\label{Heff_CrO2_sep_1nn}
\end{equation}

The second nearest neighbor:
\begin{equation}
\tilde{t}_{m',m}^{\frac{1}{2}\frac{1}{2}\frac{1}{2} ({\rm Cr1}\rightarrow{\rm Cr2})} 
{\rm =} \left(
\begin{array}{rrr}
 -26 &      &      \\
     &    0 &    0 \\
     & -202 &  135 
\end{array}
\right),
\label{Heff_CrO2_sep_2nn}
\end{equation}
where the same definitions of Eqs. 
(\ref{Heff_CrO2_basis})--(\ref{Heff_CrO2_2nn}) are used.
Since the Wannier orbitals are more extended 
the hoppings are modified slightly, but not changed significantly.
A few large hoppings of far NN are also reduced; 
$t_{xy,xy}^{002 ({\rm Cr1}\rightarrow{\rm Cr1})}=-21$ meV, 
$t_{xy,xy}^{1\bar{1}0 ({\rm Cr1}\rightarrow{\rm Cr1})}=-16$ meV, and
$t_{yz-zx,yz-zx}^{11\frac{1}{2} ({\rm Cr1}\rightarrow{\rm Cr1})}=-39$ meV.
In addition, there is no matrix elements in LDA Hamiltonian
between $d_{xy}$ and $d_{yz\pm zx}$ block, that is there is no LDA interaction
between them, by virtue of our construction of NMTO Wannier functions.

The possibly minimal set of Wannier function may provide new insights.
At the first glance one can separate the narrow $d_{xy}$ orbitals from 
the extended  $d_{yz\pm zx}$ states. The former would be treated as 
corelike (classical) spin $S=1/2$, meanwhile in the later dispersive  
$d_{yz\pm zx}$ bands the Coulomb repulsion would be treated in a usual 
quantum many-body way. Such a Kondo-lattice type model, for CrO$_2$
would be described by the following Hamiltonian:
\begin{align}
H =& \sum_{i,j, m,m^\prime \sigma} t_{i,j}^{m,m^\prime} c_{im\sigma}^\dagger c_{jm^\prime\sigma} 
\notag \\
 -& J \sum_{i,m}
\biggl[ S_{i,d_{xy}}^z(c_{im\uparrow}^\dagger c_{im\uparrow} - 
c_{im\downarrow}^\dagger c_{im\downarrow}) \notag \\
& +\frac 1 2 (S_{i,d_{xy}}^+ c_{im\downarrow}^\dagger c_{im\uparrow} + 
S_{i,d_{xy}}^- c_{im\downarrow}^\dagger c_{im\uparrow}) \biggr].
\label{kondo_ham}
\end{align}
The first term denotes the hopping of the conduction electrons between the NN sites
$i,j$ with a hoping matrix element $t_{i,j}^{m,m^\prime}$ between the $d_{yz\pm zx}$ 
orbitals, described by the $m,m^\prime$ indices. The  $d_{yz\pm zx}$ electrons and 
the localized $d_{xy}$ spins interaction is given by the second term as an exchange 
coupling ($J<0$ for antiferomagnetic and $J>0$ for ferromagnetic coupling). The 
ferromagnetic coupling between the core spins and conduction electrons, favors 
ferromagnetic ordering, because the hopping amplitudes of the conduction electrons 
reach the maximum possible values if the core 
spins are aligned. Many of the Mn-type collosal magnetoresistance materials (CMR) 
can be described by such a model. 
Recently a realistic LDA+DMFT calculation for such a model 
is carried out.~\cite{yamasaki06}
The $d$ shell of the Mn$^{3+}$ in the undoped 
antiferromagnetic insulator contains three electrons in the $t_{2g}$ orbitals forming 
a core spin of magnitude $S=3/2$, which due to strong Hund's rule coupling couples 
ferromagnetically to one additional electron in one of the $e_g$ orbitals. For a certain 
doping CMR materials are ferromagnetic metals because of additional holes in the $e_g$ 
conduction bands. 

In contrast to the manganites, in CrO$_2$, the $d_{xy}$ spin has a smaller in 
value $S=1/2$, and the splitting of the localized $d_{xy}$ orbital with repespect to the 
itinerant $d_{yz\pm zx}$ one is smaller [$\approx 370$ meV, 
see. Eq. (\ref{Heff_CrO2_sep_0nn})].

It was shown recently~\cite{Veenendaal04} that for a proper 
definition of the many-body green function in CrO$_2$, the complete
$t_{2g}$ manifold is required. Moreover a better description of polarization
can be obtained considering the competition between quasiparticle description
around the Fermi level, and a local moment behavior at higher energies, 
above the Fermi level.~\cite{Veenendaal04}

According to our results, NMTO can give a minimal model with a possibly minimal 
set of distinct Wannier $d_{xy}$ and $d_{yz\pm zx}$ functions. This result does
not exclude the possibility of integrating out the degrees of freedom connected
to the narrow/extended  $d_{xy}$/$d_{yz\pm zx}$ orbitals. Physically this would 
correspond to the dualistic character of the electronic strcutre of CrO$_2$
around the Fermi level.~\cite{Huang03}

\section{Effective Coulomb repulsion\label{effU}}
The other essential component for a model Hamiltonian describing correlations,
is the average Coulomb interaction parameter $U$. This term act on the diagonal
part of the effective Hamiltonian and corresponds to the screened electron-electron
repulsion. Aryasetiawan {\it et al.} pointed out recently~\cite{screenedU}
that a rigorous way to define this quantity can be formulated in terms of
path integrals by performing a partial trace, over the degrees of freedom,
that one wants to eliminate. However, in practice for realistic materials,
the elimination of degrees of freedom is a very difficult procedure.

In order to evaluate the average Coulomb interaction on the $d$ atoms and
the corresponding exchange interactions we start with the constrained LDA
method.~\cite{Dederichs84,Norman86,McMahan88,Gunnarsson89,Hybertsen89,AnisimovU}
In this approach the Hubbard $U$ is calculated from the total energy variation
with respect to the occupation  number of the localized orbitals. In such a scheme 
the metallic screening is rather inefficient for 3$d$ transition metals.~\cite{AnisimovU}
The perfect metallic screening will lead to a smaller value of $U$. Unfortunately, 
there are no reliable schemes to calculate $U$ in metals,~\cite{Solovyev05} therefore 
in our previous works~\cite{Chioncel03,femnsb,Chioncel05,vas} we choose some intermediate 
values of $U$ from $2$ to $4.8$ eV and $J=0.9$ eV.  

\begin{table}[b]
\caption{\label{tab2}
The constrained LDA values of the average Coulomb and exchange interactions. 
The second raw 
coresponds to the results when the $e_g$ orbitals screen the $t_{2g}$ ones. 
This type of screening would be more appropriate for the CrO$_2$ case. 
}
\begin{ruledtabular}
\begin{tabular}{cccccccccc}
&                   && \multicolumn{2}{c}{NiMnSb} & & \multicolumn{2}{c}{CrO$_2$} & \\
&                   && $U$ (eV) & $J$ (eV) && $U$ (eV) & $J$ (eV) & \\ \cline{3-9}
&$t_{2g}$ and $e_g$ && 4.80 & 0.93 && 3.50 & 0.90 & \\              
&$t_{2g}$           && 4.25 & 0.93 && 3.00 & 0.87 & \\ 
\end{tabular}
\end{ruledtabular}
\end{table}

In the case of NiMnSb the constrained LDA calculation indicates 
that the average Coulomb interaction between the Mn $3d$ electrons is 
about $U=4.8$ eV with an exchange interaction energy about $J=0.9$ eV 
as seen in Table \ref{tab2}. A reduced value of $U$ can be obtained if one 
consider that the Mn $e_{g}$ orbitals participate in the screening of 
Mn $t_{2g}$,~\cite{Solovyev96,Pickett} reducing the average Coulomb interaction 
on the Mn atoms to a value of $U=4.2$ eV. However, our DMFT 
results~\cite{Chioncel03} showed that the many-body effects are equally essential 
for $t_{2g}$ and $e_{g}$ orbitals, therefore a model with only Mn $t_{2g}$, 
even it may capture the main physical results, would suffer of incompleteness.
Note that physical results for NiMnSb are not very sensitive to the value 
of $U$, as it was demonstrated.~\cite{Chioncel03}

Concerning the local Coulomb interaction $U$, in CrO$_2$, we saw from the analysis 
of the Cr-3$d$ manifold that the higher energy $e_g$ bands makes no noticeable 
contribution to the Fermi level, however they could participate in the screening 
of the $t_{2g}$ orbitals,~\cite{Solovyev96,Pickett} giving the following values for 
$U$=3 eV and $J$=0.87 eV.

\section{Summary}
Spintronics requires the search for new materials such as half-metallic 
ferromagnets, whose properties are commanded by their electronic structure 
and electron-electron correlations. To understand the collective effects 
a clear picture of the interplay between the microscopic
interactions is necessary. 
Model Hamiltonians allows a controlled reduction of the {\it ab initio}
information into a few dominant material specific hopping and interaction 
parameters. In this paper we have applied the downfolding procedure 
within the $N$th order muffin-tin orbital (NMTO) method to obtain model Hamiltonian 
parameters for two ferromagnetic half metals NiMnSb and CrO$_2$. 

For NiMnSb, the present Wannier orbital results confirm the previous 
conclusions of Ref.~[\onlinecite{deGroot83}] based on the gamma point analysis. However 
based on the whole Brillouin zone analysis we found that $d$-$d$ hybridization 
between the transition metal atoms is essential for the gap formation 
as the $p$-$d$ hybridization previously discussed.~\cite{deGroot83}
Due to the significant Mn exchange splitting, a different hybridization takes 
place in the spin-up and spin-down channels, which determines the appearance 
of the gap only for spin-down electrons meanwhile for spin-up states a metallic 
character is evidenced.  

A downfolding calculations which would include the Sb-$p$ and Ni-$d$ 
orbitals, would increase considerably the computational demand. However, the 
physical picture would not be very much changed, because an antibonding Mn 
$d$ Wannier orbital including Sb $p$ and Ni $d$ character as a tail should 
play an important role for half-metallic nature around Fermi level.
Therefore, the effective Hamiltonian described by Eqs. 
(\ref{Heff_NiMnSb_0nn})--(\ref{Heff_NiMnSb_2nn}), 
represents a good starting point to investigate many-body effects. 

Accordingly many-body correlation effects would be of primary importance for 
the Mn orbitals the Ni one could be neglected. In our previous 
many-body results~\cite{Chioncel03,femnsb} we made use of the above conclusions, 
so correlation effects were considered only for Mn-$d$ orbitals. The LDA+DMFT 
results showed the existence of nonquasiparticle states (NQP) 
state~\cite{edwards,IK} in the minority 
spin channel. These states are associated with finite temperature spin fluctuation 
processes which play an important role in depolarization. In the majority spin 
channel, FeMnSb (Ref.~\onlinecite{femnsb}) has a larger DOS at the Fermi level in 
comparison with NiMnSb.~\cite{Chioncel03} This contributes to a stronger 
finite temperature depolarization in FeMnSb, which increase with the 
on-site Coulomb interaction. 

For CrO$_2$ the bands around the Fermi level are  primarily chromium $3d$ 
states of $t_{2g}$ manifold, with $e_g$ bands situated higher in energy by the 
crystal-field splitting. The $t_{2g}$ orbitals are further splitted into a single 
$d_{xy}$ and nearly degenerate $d_{yz \pm zx}$ bands due to the orthorhombical 
distortion of CrO$_6$ octahedra. We discussed the direct and the mediated 
Cr-$d$, O-$p$, or Cr $t_{2g}$-$e_g$ interactions. Despite the 
differences between Cr $t_{2g}$-$e_g$ orbitals their interaction play 
an important role not only in characterizing the crystal-field splitting, 
but also in the 
general picture of bonding in the rutile structure. 

The basis set of the Wannier orbitals for CrO$_2$ is formed in $t_{2g}$ manifold.
In the framework of the NMTO technique two distinct downfolding procedures are possible.

In the first approach a full $t_{2g}$ construction is possible, i.e., a single $d_{xy}$ 
and a nearly degenerate $d_{yz\pm zx}$ orbitals represents the basis set. We believe that 
such a basis set can be used in connection with the multiorbital Hubbard Hamiltonian 
described in Eq. (\ref{hub_ham}). 

A second approach is made possible due to a weak $d_{xy}$ and $d_{yz\pm zx}$ coupling. 
In this case the lowest $d_{xy}$ and the higher $d_{yz\pm zx}$ bands can be derived 
separately as shown in Fig.~\ref{bnds_nmto_cro2}. For such a situation, 
downfolded NMTO Hamiltonian with a set of those Wannier orbitals
$\{ d_{xy} \}$ and $\{ d_{yz+ zx}$, $d_{yz- zx} \}$,
is block diagonal, that is, there is no hopping
between the lowest $d_{xy}$ and the higher $d_{yz\pm zx}$ orbitals.
This fact gives a strong impression for a Kondo-lattice type model Hamiltonian 
described by Eq. (\ref{kondo_ham}). 
In a future work we report on the qualitative/quantitative 
differences of such models applied to CrO$_2$.

Similarly to our previous DMFT results on semi-Heuslers~\cite{Chioncel03,femnsb} and 
zinc-blende structures,~\cite{Chioncel05,vas} further work will include the many body 
effects in CrO$_2$ and the analysis of nature of the NQP states in these classes of 
half-metallic ferromagnets.

\begin{acknowledgments} 
We are grateful for enlightening discussions
with R.A. de Groot, O. Jepsen, O. Gunnarsson, K. Held, and E. Arrigoni. 
L.C. acknowledges financial 
support offered by the Austrian Science Foundation FWF Project No. 
P18505-N16. 
\end{acknowledgments}


\begin{thebibliography}{99}      

\bibitem{ufn}  V. Yu. Irkhin and M. I. Katsnelson,
Usp. Fiz. Nauk \textbf{164}, 705 (1994) [Phys. Usp.
\textbf{37}, 659 (1994)].

\bibitem{Wolf01} S. A. Wolf, D. D. Awschalom, R. A. Buhrman, 
J. M. Daughton, S. von Moln\'ar, M. L. Roukes, A. Y. Chtchelkanova, and
D. M. Treger,
Science {\bf 294}, 1488 (2001).

\bibitem{sarma1} I. \v{Z}uti\`{c}, J. Fabian, and S. Das Sarma, Rev. Mod.
Phys. {\bf 76}, 323 (2004).

\bibitem{deGroot83} R.A. de Groot, F.M. Mueller, P.G. van Engen, and
K.H.J. Buschow, \prl {\bf 50}, 2024 (1983).


\bibitem{Soulen98} R. J. Soulen, Jr., M. Byers, M. S. Osofsky,
B. Nadgorny, T. Ambrose, S. F. Cheng, P. R. Broussard, C. T. Tanaka,
J. Nowak, J. S. Moodera, A. Barry, and J. M. D. Coey, 
Science {\bf 282}, 85 (1998);
R. J. Soulen, Jr., M. S. Osofsky, B. Nadgorny, T. Ambrose, P. Broussard,
S. F. Cheng, J. M. Byers, C. T. Tanaka, J. Nowack, J. S. Moodera,
G. Laprade, A. Barry, and M. D. Coey, J. Appl. Phys. {\bf 85}, 4589 (1999).

\bibitem{Clowes04} S. K. Clowes, Y. Miyoshi, Y. Bugoslavsky,
W. R. Branford, C. Grigorescu, S. A. Manea, O. Monnereau, and
L. F. Cohen, \prb {\bf 69}, 214425 (2004).

\bibitem{Zhu01} W. Zhu, B. Sinkovic, E. Vescovo, C. Tanaka,
and J. S. Moodera, \prb {\bf 64}, 060403(R) (2001).

\bibitem{Sicot06} M. Sicot, P. Turban, S. Andrieu, A. Tagliaferri,
C. De Nadai, N. B. Brookes, F. Bertran, and F. Fortuna,
J. Magn. Magn. Mater. {\bf 303}, 54 (2006).
%

\bibitem{Wijs01}G. A. de Wijs and R. A. de Groot, 
Phys. Rev. B {\bf 64}, 020402(R) (2001).

\bibitem{Ji01} 
Y. Ji, G. J. Strijkers, F. Y. Yang, C. L. Chien, J. M. Byers, 
A. Anguelouch, G. Xiao, and A. Gupta,
\prl {\bf 86}, 5585 (2001).

\bibitem{Parker01}
J. S. Parker, S. M. Watts, P. G. Ivanov, and P. Xiong, 
\prl {\bf 88}, 196601 (2002).

\bibitem{Dedkov02} Y.S. Dedkov, M. Fonine, C. Konig, U. Rudiger, 
G. Guntherodt, S. Senz, and D. Hesse, 
Appl. Phys. Lett. {\bf 80}, 4181 (2002). 

\bibitem{Coey02} J.M.D. Coey, J. Versluijs, and M. Venkatesen,
J. Phys. D {\bf 35}, 2457 (2002).

%
\bibitem{Keizer06}
R. S. Keizer, S. T. B. Goennenwein, T. M. Klapwijk, G. Miao, G. Xiao, 
and A. Gupta,
Nature (London) {\bf 439}, 825 (2006).
%

\bibitem{ourDMFT} A. I. Lichtenstein and M. I. Katsnelson, Phys. Rev. B {\bf 57},
6884 (1998).

\bibitem {DMFT1}
W. Metzner and D. Vollhardt,
Phys. Rev. Lett. {\bf 62}, 324 (1989). 

\bibitem {DMFT2}
A.Georges, G. Kotliar, W. Krauth, and M. Rozenberg,
Rev. Mod. Phys. {\bf 68}, 13 (1996).

\bibitem{anisDMFT} V. I. Anisimov, A. I. Poteryaev, M. A. Korotin, A. O.
Anokhin, and G. Kotliar, J. Phys.: Condens. Matter {\bf 9}, 7359
(1997).

\bibitem{Katsnelson99} M. I. Katsnelson and A. I. Lichtenstein, J. Phys.:
Condens. Matter {\bf 11}, 1037 (1999).

\bibitem{HeldLDADMFT}
K. Held, 
I. A. Nekrasov, G. Keller, V. Eyert, N. Bl\"umer, A. K. McMahan,
R. T. Scalettar, Th. Pruschke, V. I. Anisimov, and D. Vollhart,
Psi-k Newsletter \#56, 65 (2003)
[http://psi-k.dl.ac.uk/newsletters/News\_56/Highlight\_56.pdf].

\bibitem{EMTODMFT} L. Chioncel, L.Vitos, I. A. Abrikosov, J. Koll\'ar,
M. I. Katsnelson, and A. I. Lichtenstein, Phys. Rev. B {\bf 67},
235106 (2003).

\bibitem{today} G. Kotliar and D. Vollhardt, Phys. Today {\bf 57}, 
53 (2004);  G. Kotliar, S. Y. Savrasov, K. Haule, V. S. Oudovenko,
O. Parcollet, and C.A. Marianetti, to be published in Rev. Mod. Phys. 
{\bf 78} (2006), cond-mat/0511085.

\bibitem{Chioncel03} L. Chioncel, M. I. Katsnelson, R. A. de Groot, and
A. I. Lichtenstein, Phys. Rev. B {\bf 68}, 144425 (2003).

\bibitem{femnsb}L. Chioncel, E. Arrigoni, M.I. Katsnelson, 
and A.I. Lichtenstein,
\prl {\bf 96}, 137203 (2006).

\bibitem{Chioncel05} L. Chioncel, M. I. Katsnelson, G. A. de Wijs, R. A. de
Groot, and A. I. Lichtenstein, Phys. Rev. B {\bf 71}, 085111
(2005).

\bibitem{vas}L. Chioncel, Ph. Mavropoulos, M. Le\v{z}ai\'c, S. Bl\"ugel, 
E. Arrigoni, M.I. Katsnelson, and A.I. Lichtenstein, 
\prl {\bf 96}, 197203 (2006).

\bibitem{edwards} D. M. Edwards and J. A. Hertz, J. Phys. F {\bf 3}, 2191 (1973).

\bibitem{IK} V. Yu. Irkhin and M. I. Katsnelson, Fizika Tverdogo Tela
{\bf 25}, 3383 (1983) [Sov. Phys. - Solid State {\bf 25}, 1947
(1983)]; J. Phys. : Condens. Matter {\bf 2}, 7151 (1990).

\bibitem{Sanyal}B. Sanyal, L. Bergqvist, and O. Eriksson, Phys. Rev. B {\bf 68}, 054417 (2003).

\bibitem{andersen00} O. K. Andersen and T. Saha-Dasgupta,
\prb {\bf 62}, R16219 (2000);
O. K. Andersen, O. Jepsen, and G. Krier,
in {\it Methods of Electronic Structure Calculations}, 
edited by V. Kumar, O. K. Andersen, and A. Mookerjee 
(World Scientific, Singapore, 1994), pp. 63--124;
O. K. Andersen, T. Saha-Dasgupta, R. W. Tank, C. Arcangeli, O. Jepsen, and G. Krier, 
in {\it Electronic Structure and Physical Properties of Solids. 
The Uses of the LMTO Method}, edited by H. Dreysse, 
Springer Lecture Notes in Physics (Springer, New York, 2000), pp. 3--84;
O. K. Andersen, T. Saha-Dasgupta, S. Ezhov, L. Tsetseris, 
O. Jepsen, R. W. Tank, C. Arcangeli, and G. Krier,
Psi-k Newsletter \#45, 86 (2001)
[http://psi-k.dl.ac.uk/newsletters/News\_45/Highlight\_45.pdf].

\bibitem{zurek05} E. Zurek, O. Jepsen, and O. K. Andersen,
ChemPhysChem {\bf 6}, 1934 (2005).

\bibitem{lmto} O. K. Andersen,
\prb {\bf 12}, 3060 (1975).

\bibitem{tblmto} O. K. Andersen and O. Jepsen,
\prl {\bf 53}, 2571 (1984).

\bibitem{tb47} The Stuttgart TB-LMTO-ASA code, version 4.7.
See: http://www.fkf.mpg.de/andersen/.

\bibitem{PavariniNJP} E. Pavarini, A. Yamasaki, J. Nuss, and O. K. Andersen,
New J. Phys. {\bf 7}, 188 (2005).

\bibitem{Ogut}  S. \"{O}\u{g}\"{u}t and K. M. Rabe, \prb {\bf 51}, 10443 (1995).

\bibitem{Galanakis}  I. Galanakis, P. H. Dederichs, and N. Papanikolaou,
\prb {\bf 66}, 134428 (2002).

\bibitem{Nanda} B. R. K. Nanda and I. Dasgupta,
J. Phys.: Condens. Matter {\bf 15}, 7307 (2003).

\bibitem{Kulatov} E. Kulatov and I. I. Mazin,
J. Phys.: Condens. Matter {\bf 2}, 343 (1990).

\bibitem{CrO2str} B.J. Thamer, R.M. Douglass, and E. Staritzky, J. Am. Chem. Soc.
{\bf 79}, 547 (1957).

\bibitem{Lewis97} S.P. Lewis, P.B. Allen and T. Sasaki, Phys. Rev. B {\bf 55},
10253 (1997).

\bibitem{Korotin98} M.A. Korotin, V.I. Anisimov, D.I. Khomskii,
and G.A. Sawatzky,
\prl {\bf 80}, 4305 (1998).

\bibitem{Schlottmann03} P. Schlottmann,
\prb {\bf 67}, 174419 (2003).

\bibitem{Chamberland77} B.L. Chamberland, CRC Crit. Rev. Solid State Mater. 
Sci. {\bf 7}, 1 (1977).

\bibitem{Tsujioka97} T. Tsujioka, 
T. Mizokawa, J. Okamoto, A. Fujimori, M. Nohara, H. Takagi, K. Yamaura,
and M. Takano,
\prb {\bf 56}, R15509 (1997).

\bibitem{Stagarescu00} C.B. Stagarescu, 
X. Su, D. E. Eastman, K. N. Altmann, F. J. Himpsel, and A. Gupta,
\prb {\bf 61}, R9233 (2000).

\bibitem{Suzuki98} K. Suzuki and P. M. Tedrow,
\prb {\bf 58}, 11597 (1998).

\bibitem{Singley99} E.J. Singley, 
C. P. Weber, D. N. Basov, A. Barry, and J. M. D. Coey,
\prb {\bf 60}, 4126 (1999).

\bibitem{Schwarz86} K.H. Schwarz, J. Phys. F
{\bf 16}, L211 (1986).

\bibitem{Mazin99} I.I. Mazin, D.J. Singh, and C. Ambrosch-Draxl,
\prb {\bf 59}, 411 (1999).

\bibitem{Hartmann03} M.S. Laad, L. Craco, and E. Muller-Hartmann,
\prb {\bf 64}, 214421 (2001);  L. Craco, M.S. Laad, and E. Muller-Hartmann,
\prl {\bf 90}, 237203 (2003).

\bibitem{Sorantin92} P. I. Sorantin and K. Schwarz,
Inorg. Chem. {\bf 31}, 567 (1992).

\bibitem{note} Note that the local coordinate system in the paper
Ref.~[\onlinecite{Sorantin92}] is different from the one presented in
Ref.~[\onlinecite{Korotin98}]. In the present paper we use the convention 
presented in the later Ref.~[\onlinecite{Korotin98}].

\bibitem{CrO2str2} P. Porta, M. Marezio, J. P. Remeika, and P. D. Dernier,
Mater. Res. Bull. {\bf 7}, 157 (1972).

\bibitem{V2O3} T. Saha-Dasgupta, O. K. Andersen, J. Nuss,
A. Poteryaev, A. I. Lichtenstein, and A. Georges
(unpublished).

\bibitem{NaCoO2} O. K. Andersen, I. I. Mazin, O. Jepsen, M. D. Johannes,
and A. Yamasaki (unpublished).

\bibitem{yamasaki06} A. Yamasaki, M. Feldbacher, Y. -F. Yang, O. K. Andersen, 
and K. Held, 
\prl {\bf 96}, 166401 (2006).

\bibitem{Veenendaal04}M. van Veenendaal and A. J. Fedro, \prb {\bf 70},
012412 (2004).

\bibitem{Huang03}
D. J. Huang, L. H. Tjeng, J. Chen, C. F. Chang, W. P. Wu, S. C. Chung, 
A. Tanaka, G. Y. Guo, H.-J. Lin, S. G. Shyu, C. C. Wu, and C. T. Chen,
Phys. Rev. B {\bf 67}, 214419 (2003).

\bibitem{screenedU}F. Aryasetiawan, M. Imada, A. Georges, G. Kotliar, S. Biermann, and A. I. Lichtenstein,
Phys. Rev. B {\bf 70}, 195104 (2004).

\bibitem{Dederichs84} P. H. Dederichs, S. Bl\"ugel, R. Zeller, and H. Akai,
Phys. Rev. Lett. {\bf 53}, 2512 (1984).

\bibitem{Norman86} M. R. Norman and A. J. Freeman,
Phys. Rev. B {\bf 33}, R8896 (1986).

\bibitem{McMahan88} A. K. McMahan, R. M. Martin, and S. Satpathy,
Phys. Rev. B {\bf 38}, 6650 (1988).

\bibitem{Gunnarsson89} O. Gunnarsson, O. K. Andersen, O. Jepsen, and J. Zaanen,
Phys. Rev. B {\bf 39}, 1708 (1989).

\bibitem{Hybertsen89} M. S. Hybertsen, M. Schl\"uter, and N. E. Christensen,
Phys. Rev. B {\bf 39}, 9028 (1989).

\bibitem{AnisimovU} V. I. Anisimov and O. Gunnarsson, Phys. Rev. B {\bf 43}, 7570 (1991).

\bibitem{Solovyev05}For recent progress, see Ref.~[\onlinecite{screenedU}];
I. V. Solovyev and M. Imada, Phys. Rev. B {\bf 71}, 045103 (2005); 
F. Aryasetiawan, K. Karlsson, O. Jepsen, and U. Sch\"onberger,
cond-mat/0603138.

\bibitem{Solovyev96} I. Solovyev, N. Hamada, and K. Terakura,
Phys. Rev. B {\bf 53}, 7158 (1996).

\bibitem{Pickett} W. E. Pickett, S. C. Erwin, and E. C. Ethridge, 
\prb {\bf 58}, 1201 (1998).

\end{thebibliography}
\end{document}